\documentclass[twocolumn,pra,superscriptaddress]{revtex4-1}
\usepackage{amsmath}
\usepackage{amssymb}
\usepackage{graphicx}
\usepackage{float}
\newcommand{\lgmode}[2]{LG$_{#1}^{#2}$}
\newcommand{\reffig}[1]{Figure~\ref{#1}}
\newcommand{\refeq}[1]{Equation~(\ref{#1})}

\newcommand{\hoground}{\epsilon_{ho}}
\newcommand{\holength}{a_{ho}}

\begin{document}
\title{Bose-Einstein condensation transition studies for atoms 
		confined in Laguerre-Gaussian laser modes}
\author{T.~G.~Akin}
\author{Sharon Kennedy}
\affiliation{Homer L. Dodge Department of Physics and Astronomy, University of Oklahoma, Norman, OK, USA}
\author{Ben Dribus}
\affiliation{Louisiana State University, Baton Rouge, LA, USA}
\author{Jeremy L. Marzuola}
\affiliation{Mathematics Department, University of North Carolina - Chapel Hill, Chapel Hill, NC, USA}
\author{Lise Johnson}
\affiliation{Department of Neurological Surgery, University of Washington, Seattle, WA, USA}
\author{Jason Alexander}
\affiliation{Cold Atom Optics Group, U.S. Army Research Laboratory, Adelphi, MD, USA}
\author{E.~R.~I.~Abraham}
\email[Electronic address: ]{abraham@nhn.ou.edu}
\affiliation{Homer L. Dodge Department of Physics and Astronomy, University of Oklahoma, Norman, OK, USA}
\date{\today}
\begin{abstract}
	Multiply-connected traps for cold, neutral atoms fix vortex cores 
	of quantum gases.  Laguerre-Gaussian laser modes are ideal for 
	such traps due to their phase stability.  We report theoretical 
	calculations of the Bose-Einstein condensation transition 
	properties and thermal characteristics of neutral atoms 
	trapped in multiply connected geometries formed by 
	Laguerre-Gaussian (\lgmode{p}{l}) beams.  Specifically, 
	we consider atoms confined to the anti-node of a \lgmode{0}{1} 
	laser mode detuned to the red of an atomic resonance frequency, 
	and those confined in the node of a blue-detuned \lgmode{1}{1} 
	beam.  We compare the results of using the full potential to those 
	approximating the potential minimum with a simple harmonic 
	oscillator potential.  We find that deviations between 
	calculations of the full potential and the simple harmonic 
	oscillator can be up to $3\%-8\%$ for trap parameters consistent 
	with typical experiments.
	%
	%\pacs{}
	%
\end{abstract}
\maketitle
\begin{figure}
	\begin{center}	
		\includegraphics[width=8.5cm]{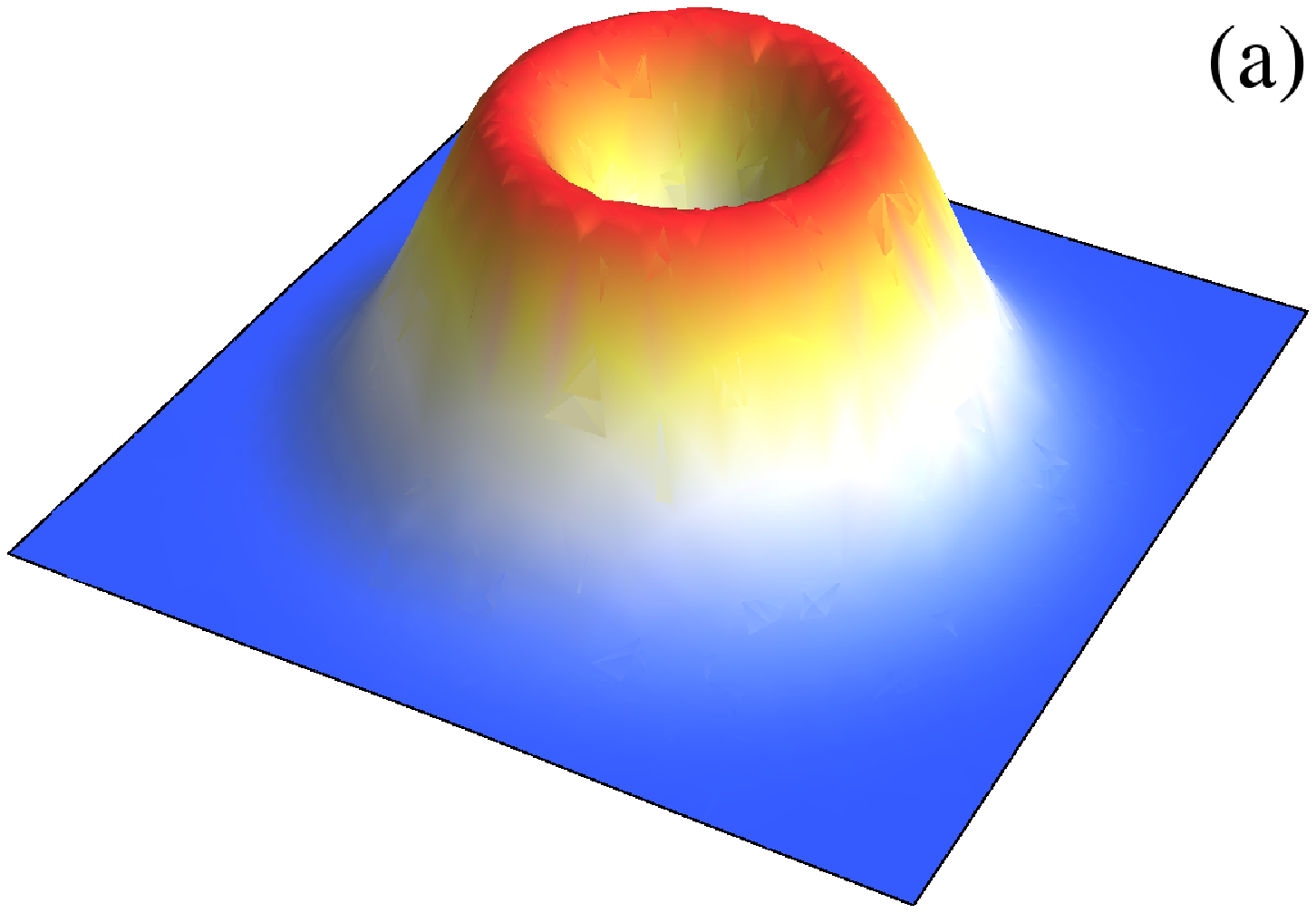}
		\includegraphics[width=8.5cm]{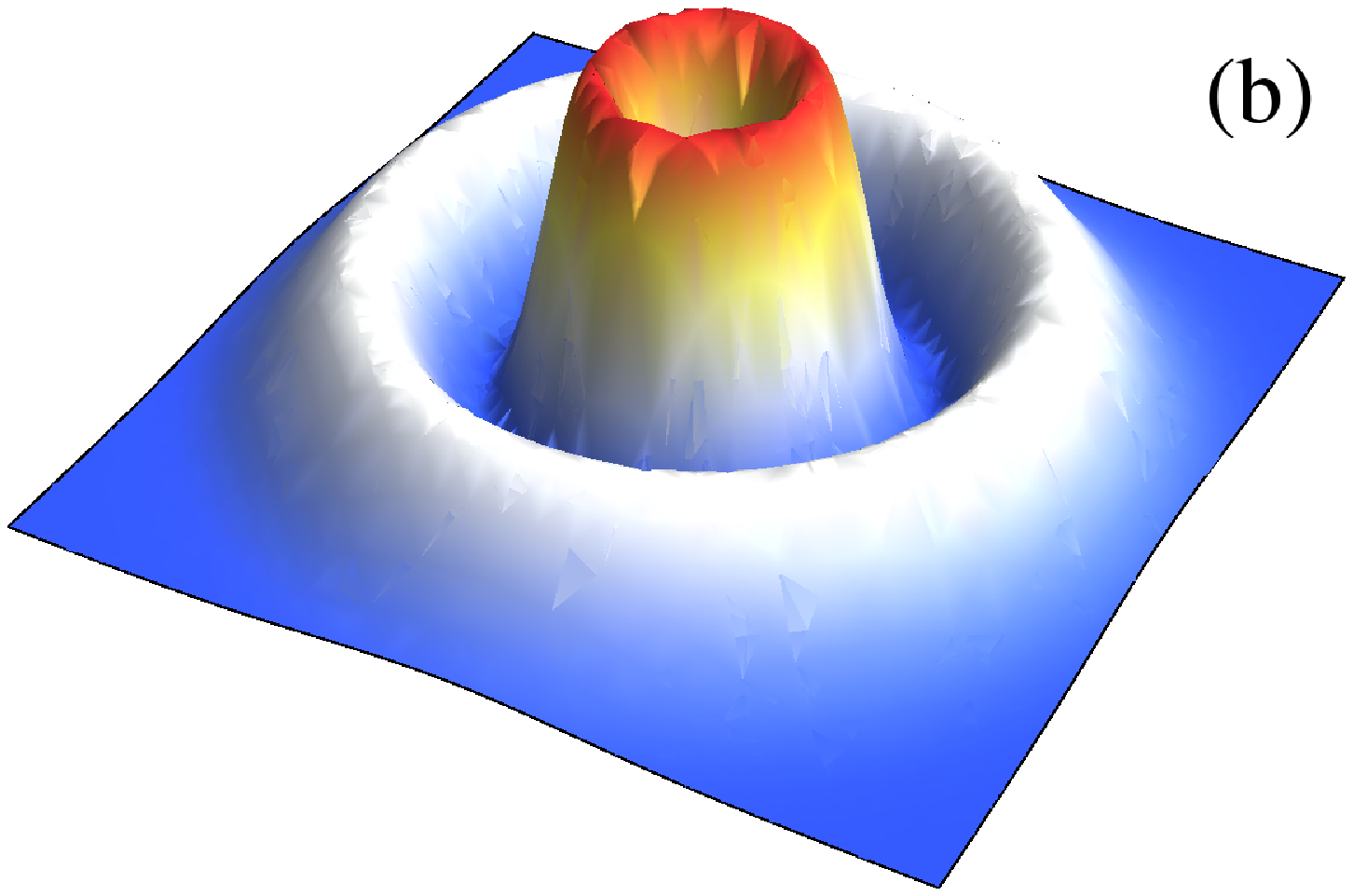}
		\caption{(Color online) The transverse intensity profile for 
		(a) the \lgmode{0}{1} mode and (b) the \lgmode{1}{1} mode.}
		\label{beamprofile}
	\end{center}
\end{figure}
\begin{figure}
	\begin{center}
		\includegraphics[width=8.5cm]{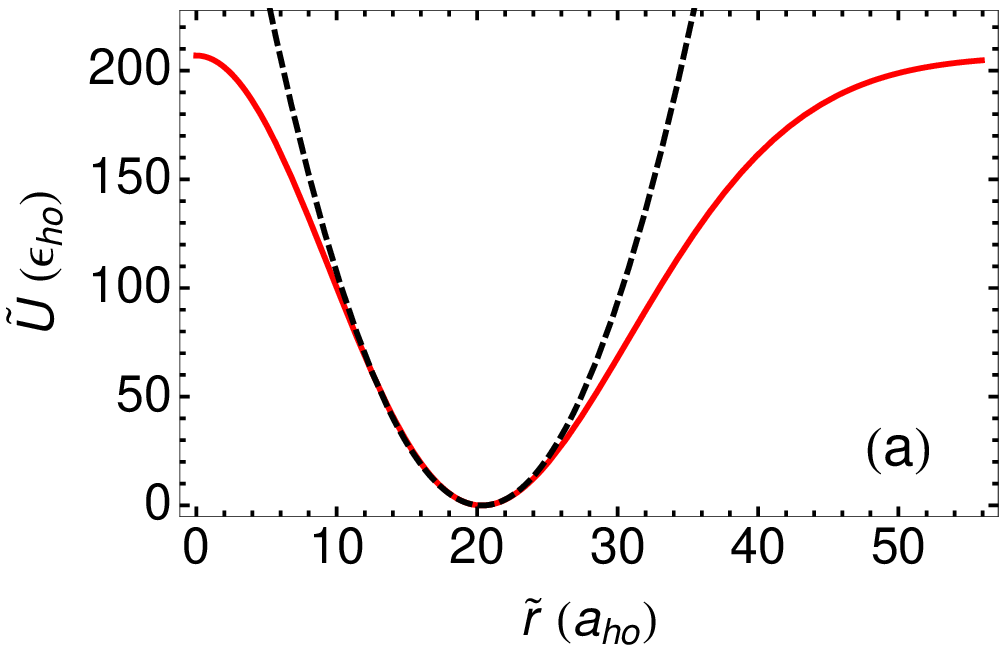}
		\includegraphics[width=8.5cm]{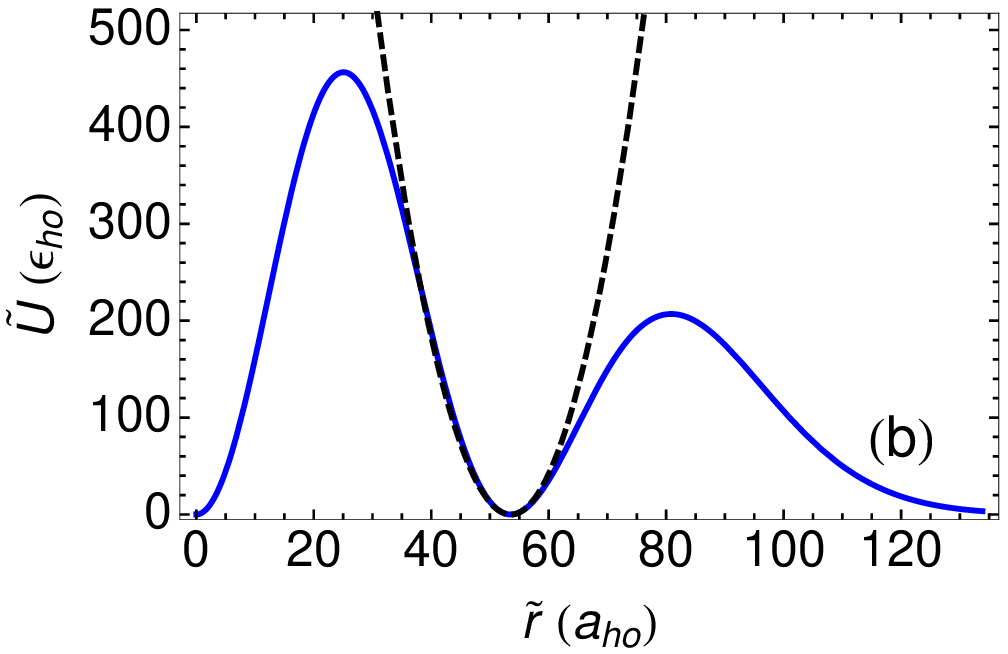}
		\caption{(Color online) The radial potential energy (in units of 
		$\hoground$) of an atom in specific \lgmode{p}{l} laser modes as 
		a function of distance from the center of the laser beam (in 
		units of $\holength$).  In both plots, the solid line 
		corresponds to the full \lgmode{p}{l} dipole potential, and the 
		dashed line represents the second order expansion about the 
		minima of the potential (SHO approximation).}
		\label{rpotential}
	\end{center}
\end{figure}
\begin{figure}
	\begin{center}
		\includegraphics[width=8.5cm]{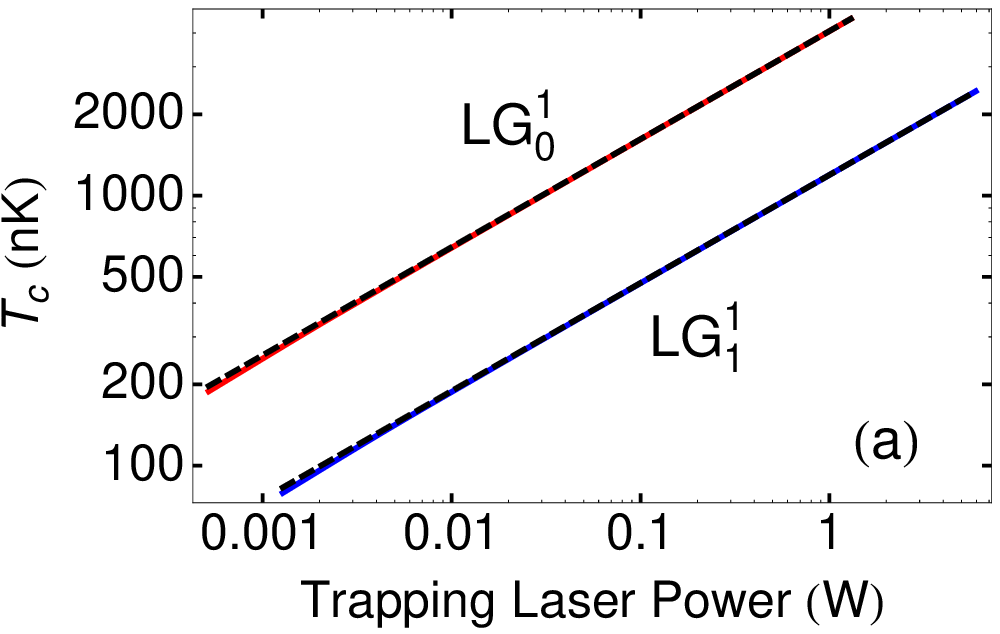}
		\includegraphics[width=8.5cm]{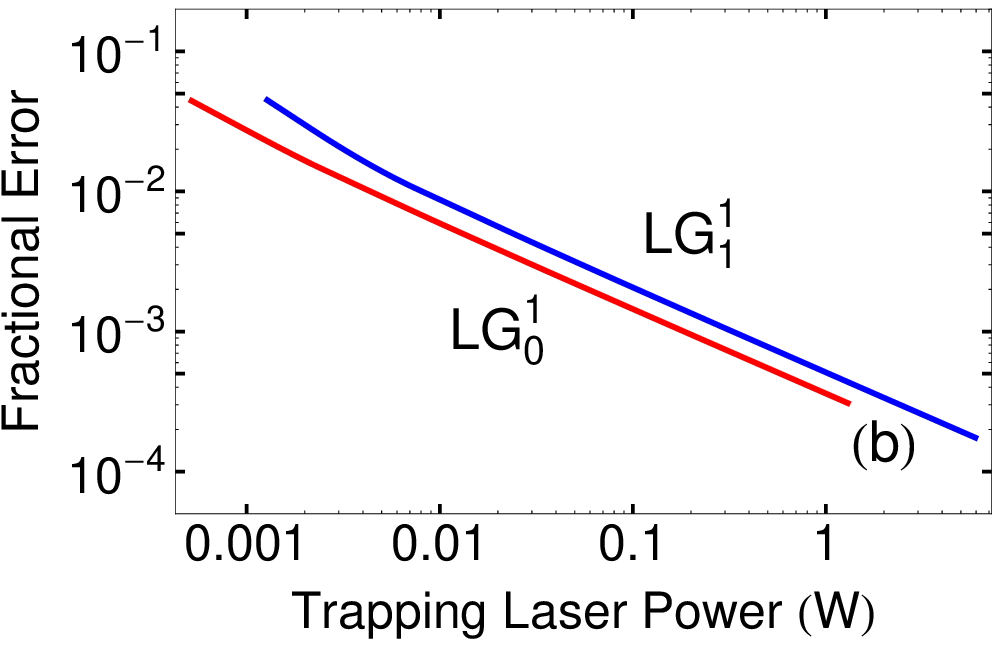}
		\caption{ (Color online) (a) $T_c$ as a function of 
		the laser power.  The beam parameters are $w_r = 50~\mu\textrm{m}$ 
		and $\Delta/2\pi =1000$~GHz.  The comparison is made using both 
		the full potential (solid curve) and the SHO approximation (dashed 
		curve) while keeping the trapping beam waist fixed and simultaneously 
		increasing the laser power.  In addition, the axial harmonic trapping 
		frequency is adjusted such that the trap remains symmetric at all 
		laser powers.  (b) The fractional error of the SHO calculation.}
		\label{tcfreq}
	\end{center}
\end{figure}
\begin{figure}
	\begin{center}
		\includegraphics[width=8.5cm]{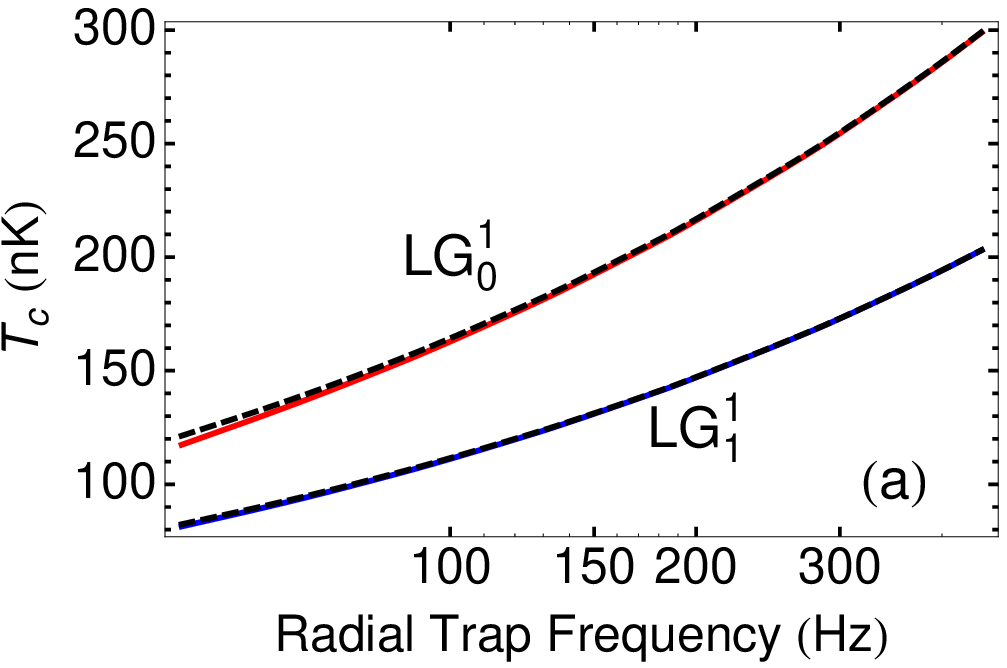}
		\includegraphics[width=8.5cm]{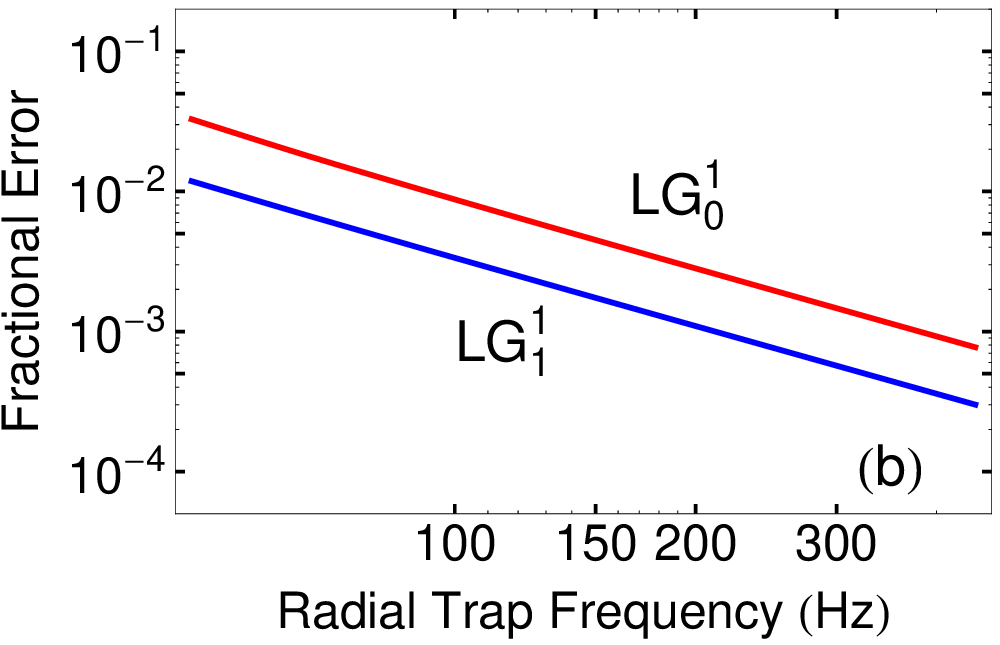}
		\caption{(Color online) (a) $T_c$ 
		calculations using both the full potential (solid curve) and the 
		SHO approximation (dashed curve) while keeping the 
		trapping beam waist fixed and 
		simultaneously increasing the laser power.  Thus, the 
		effects of adjusting the radial trapping frequency with 
		respect to the axial frequency.  Again, the axial frequency is 
		fixed at $100 \textrm{Hz}$.  
		(b) The factional error of the SHO calculation.}
		\label{tcasym}
	\end{center}
\end{figure}
\begin{figure}
	\begin{center}
		\includegraphics[width=8.5cm]{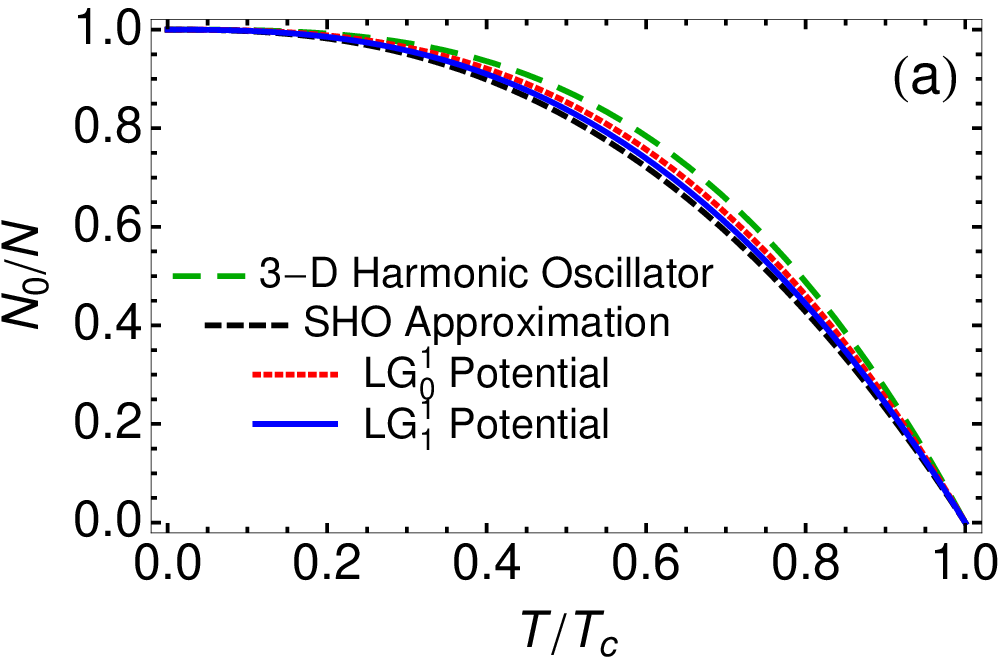}
		\includegraphics[width=8.5cm]{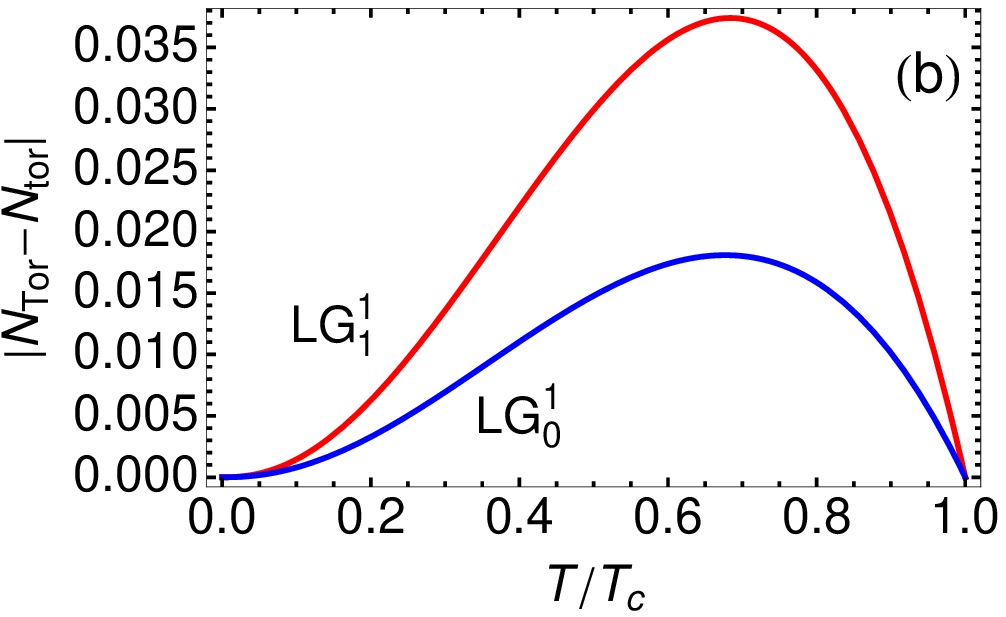}
		\caption{(Color online) (a) Plot of the fraction of particles in 
		the ground state as a function of the scaled temperature.  
		Plotted are the SHO approximation to the LG potentials (black 
		dashed), the \lgmode{1}{1} mode (blue solid), the \lgmode{0}{1} 
		mode (red dot-dashed), and the 3-D isotropic harmonic oscillator 
		(green dotted).  (b) The absolute difference in \lgmode{p}{l} fractions and the 
		fraction due to the SHO approximation}
		\label{groundfrac}
	\end{center}
\end{figure}
\begin{figure}
	\begin{center}
		\includegraphics[width=8.5cm]{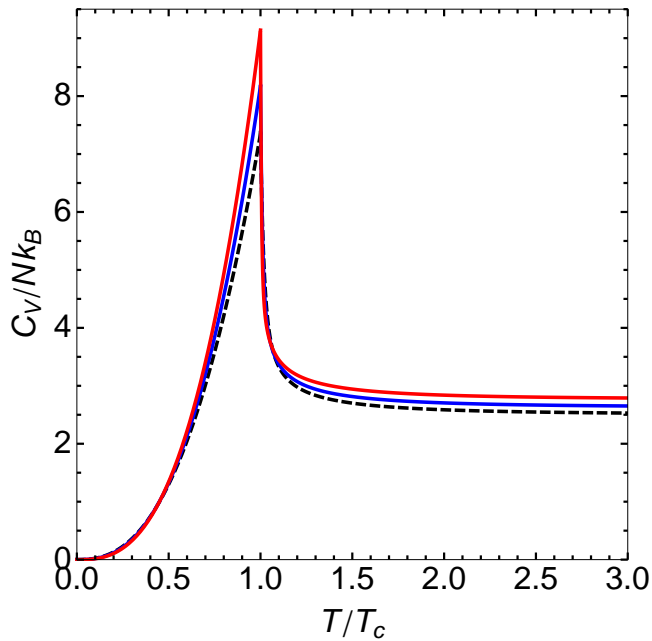}
		\caption{(Color online) Plot of the heat capacity per particle in 
		units of $k_b$ for the \lgmode{0}{1} (red), the 
		\lgmode{1}{1} (blue), and the harmonic approximation (black).  The 
		heat capacity is a function of the temperature, 
		where~$T$~is in units of~$T_c$.  The thermodynamic discontinuity 
		at~$T=T_c$~indicates a phase transition.}
		\label{heatcapacity}
	\end{center}
\end{figure}

Bose-Einstein condensation (BEC) has led to a wealth of new physics, from 
applications in inertial measurements to fundamental studies of statistical 
mechanical phenomena and superfluidity~\cite{dg99}.  Among the first 
experiments probing BEC were those that explored the thermal properties 
and transition characteristics, with particular emphasis on the effect 
of trap geometry~\cite{mav96,ejm96}.  An important avenue of 
investigation in the connection between condensed matter (super-fluidity) 
and BEC~\cite{l38} of atomic gases is the 
experimental~\cite{mah99,mcw00,arv01} and the 
theoretical~\cite{dcl98,cj09} studies of 
vortices.  It was recognized early that a trap potential with the 
ability to pin a vortex core would be an advance in the study and 
application of these quantized rotations in BEC~\cite{rac07}.  Toward 
this end, researchers have developed novel multiply-connected trap 
geometries.  A toroidal potential locally fixes a vortex state, thus 
encouraging rotational stability for more precise measurements and may 
help facilitate the development of devices utilizing vortices, such as the 
atomic squid detector~\cite{adw03} and the ultra-stable 
gyroscope~\cite{tkd09}.  Recent progress in toroidal potentials has been 
seen in the confinement of cold atoms~\cite{bsc01}, BEC~\cite{orv00,ar02}, 
and creation of vortex states~\cite{rac07}. \\
\indent An important method for creating toroidal geometries uses 
higher-order Laguerre-Gaussian laser modes.  
The Laguerre-Gaussian beams, as a set of solutions to 
Maxwell's equations \cite{abs92}, represent stable modes of laser 
propagation.  The radial electric field is proportional to an associated 
Laguerre polynomial, $L_{p}^{l}$, and a Gaussian function.  
The electric field of \lgmode{p}{l} laser modes has an azimuthal 
winding phase given by \mbox{$e^{-i l \phi}$} \cite{abs92}, 
and the $p+1$ radial intensity nodes provide a variety of 
multiply connected geometries for vortex studies.  
\lgmode{p}{l} photons have a quantized orbital angular 
momentum (OAM) of $l \hbar$ per photon.  This 
property can also be exploited to create vortices through the coupling of 
photon-matter OAM, as proposed by \cite{mzw97,kwm07} and 
demonstrated by \cite{rac07,wlb08}. \\
\indent We explore the thermal properties (the transition temperature, the population 
of atoms in the ground state, and the specific heat) of a Bose gas trapped in a \lgmode{p}{l} 
dipole trap, utilizing the complete \lgmode{p}{l} potential.  These results 
are compared to those where the confining potential minimum has been approximated as a 
simple harmonic oscillator.  Previously, the simple harmonic 
oscillator approximation has been used in calculations 
to determine the ground state energies, density profiles, and the 
transition temperature in LG beams \cite{spr99,wad00,tw01}. \\
\indent Consider an atom in an inhomogeneous laser field whose 
angular frequency, $\omega$, is near enough to a resonant 
angular frequency, $\omega_0$, that the coupling between 
any other pair of states can be ignored.  The laser  
detuning, $\Delta = \omega - \omega_0$, is sufficiently large 
that the probability of photon absorption is negligible.  The atom 
experiences an attractive force toward regions of high laser intensity 
when the detuning is negative ($\Delta<0$) and a repulsive force away from regions of 
high laser intensity when the detuning is positive ($\Delta>0$).  \\
\indent We investigate two physically realizable toroidal trapping 
geometries formed by the lowest two orders in a~\lgmode{p}{l} beam 
(with non-zero angular momentum): the \lgmode{0}{1} 
red-detuned trap and the \lgmode{1}{1} blue-detuned trap.
The intensity profile for the \lgmode{0}{1} mode is given in 
\reffig{beamprofile}~(a).  For $\Delta<0$, this shape satisfies 
the desired toroidal geometry with atoms confined to regions of high 
laser intensity.  Blue detuned dipole traps are 
preferred for applications where external 
perturbation needs to be minimized.  Atoms spend the majority of their 
time in regions of lowest intensity.  The \lgmode{1}{1} beam satisfies this 
condition, and a sample profile is shown in \reffig{beamprofile}~(b).  \\
\indent The potential energy of an atom in the presence of a \lgmode{p}{l} 
mode is given by~\cite{wad00},
\begin{eqnarray}
	\label{lgdipole}
	U_p^l=\frac{\hbar \Gamma^2}{8 I_{sat} \Delta} \frac{2 p !}{p+l} \frac{P_0}{\pi w^2} 
		\left( \frac{2r^{2}}{w^{2}} \right)^{l} \left[ L_p^l\left(\frac{2 r^2}{w^2}\right) \right]^{2} e^{-2 r^2/w^2},
\end{eqnarray}
where~$P_0$~is the total laser power, 
$\Gamma=2\pi\times 6.1$~MHz is the natural line width for~$^{87}$Rb, and $w$ is the beam 
waist.  The resonant saturation intensity is $I_{sat} = \pi h c / 3 \lambda^3 \tau$, where 
$\lambda$ is the resonance wavelength and $\tau$ is the lifetime of the resonant state.  
We work in cylindrical coordinates where the dipole potential 
only confines in the transverse $r$-dimension.  To confine along the direction of laser propagation
($z$-dimension), we will assume a harmonic potential.  One dimensional confinement can be 
constructed via an anisotropic magnetic trap whose trapping potential in the radial dimension is small enough 
to be ignored compared to the laser potential~\cite{pae95,gvl01}.   With the 
inclusion of the 1-D harmonic confinement together with the radial dipole potential in~\refeq{lgdipole}, 
the confining potential for both laser modes has the form,
\begin{eqnarray}
	\label{fullpot}
	\begin{array}{lll}
	U_0^1&=& 2 V \left( \frac{2 r^2}{w^2} \right)  e^{-2 r^2/w^2} +\frac{1}{2} m \omega^2_z z^2 \\
	U_1^1&=&V \left( \frac{2 r^2}{w^2} \right) \left( 2- \frac{2 r^2}{w^2} \right)^2 e^{-2 r^2/w^2} + 
		\frac{1}{2} m \omega^2_z z^2,
	\end{array}
\end{eqnarray}
where~$\omega_z$~is the trapping frequency in the~$z$-direction and $m$ is the mass of the confined atom.  
The constant~$V$~is the product of three physical quantities,
\begin{eqnarray*}
	V = \frac{\hbar\Gamma}{8} \frac{I_0}{I_{sat}} \frac{\Gamma}{\Delta}.
\end{eqnarray*}
The first factor is an energy scale set by the energy width of the excited state.  The second factor 
is the ratio of an intensity term ($I_0=P_0/\pi w^2$) to the saturation intensity.  
The last factor is a ratio of the line width to the laser detuning.  These factors 
combine to set the scale of the trap strength.\\
\indent The 1-D radial cross-section of both potentials in \refeq{fullpot} are plotted in 
\reffig{rpotential}.  The calculations are done in dimensionless units.  All the lengths are scaled 
by~$\holength=\sqrt{\hbar/m\omega_z}$, the harmonic oscillator length of the simple harmonic confining 
potential in the~$z$-direction.  All the energies are scaled by~$\hoground =\hbar \omega_z/2$, 
the energy of the corresponding 1-D ground state in the~$z$-direction.  Dimensionless quantities 
are identified with a tilde ($\sim$) set above the term:~$\tilde{r}=r/\holength$, 
$\tilde{z}=z/\holength$,~$\tilde{w} = w/\holength$,~$\tilde{\epsilon}=\epsilon/\hoground$, 
and~$\tilde{v}=V/\hoground$.  Making these substitutions into the potentials,~\refeq{fullpot} becomes,
\begin{eqnarray}
	\label{unitlesspot}
	\begin{array}{lll}
	\tilde{U}_0^1 &=& - 4 \frac{\tilde{v}}{\tilde{w}^2} \tilde{r}^2 e^{-2\tilde{r}^2/\tilde{w}^2} + \tilde{z}^2 \\
	\tilde{U}_1^1 &=& 8 \frac{\tilde{v}}{\tilde{w}^6}\tilde{r}^2 \left(\tilde{w}^2-\tilde{r}^2\right)^2
		e^{-2\tilde{r}^2/\tilde{w}^2}+\tilde{z}^2. 
	\end{array}
\end{eqnarray}
%
%--------------------------------------------------------------------------
%
\indent We explore a few of the thermal properties of a cold Bose gas 
confined in a toroidal potential.  One such property is the BEC 
transition temperature,~$T_c$.  We calculate~$T_c$~following the 
procedure outlined in~\cite{bpk87}.  In the limit when~$N \rightarrow \infty$~and~$k_B T$~is 
much larger than the energy spacing of the confining potential, then
\begin{eqnarray}
	\label{npart}
	N = N_0 + \int^{\infty}_0 \rho (\epsilon ) n (\epsilon ) d\epsilon,
\end{eqnarray}
where~$N_0$~is the number of atoms in the ground state,~$\rho (\epsilon)$~is the density of states, 
and~$n(\epsilon)$~is the usual Bose distribution occupation number \mbox{$1/(e^{\beta(\epsilon-\mu)}-1)$}, 
where~$\beta = 1/k_B T$~and~$\mu$~is the chemical potential.  The density of states is
\begin{eqnarray}
	\label{denofstates}
	\rho (\epsilon) = \frac{(2m)^{3/2}}{(2\pi)^2\hbar^3} \int_{V^*(\epsilon)} 
		\sqrt{\epsilon - U(\textbf{r})}d\textbf{r},
\end{eqnarray}
where~$V^*(\epsilon)$~is the classical spatial volume spanned by a particle with energy~$\epsilon$. \\
\indent As the phase space of the system decreases, the ground state population 
remains unoccupied until the chemical potential approaches zero.  Thus, 
the transition temperature,~$T_c$, can be found by setting~$\mu=0$~and 
$N_0=0$.  The heat capacity,~$C(T)$,~is piecewise continuous, but a discontinuity at~$T_c$ 
is a signature of the BEC phase transition.  The heat capacity 
is~$C(T)=\partial E(T)/\partial T$~where the total energy of the system is
\begin{eqnarray}
	\label{energy}
	E(T)=\int^{\infty}_{0} \epsilon \rho ( \epsilon ) n ( \epsilon ) d\epsilon.
\end{eqnarray}
\indent Equations (\ref{npart}) through (\ref{energy}) can be solved 
analytically when the potential is expanded up to the second order 
about the potential minimum.  This is the simple harmonic oscillation 
(SHO) approximation used in~\cite{spr99,wad00,tw01}.  
For our system and notation, the confining potentials are given by
\begin{eqnarray}
	\label{unitlesstor}
	\begin{array}{lll}
	(\tilde{U}_{0}^{1})_{ho} &=& \frac{8\tilde{v}}{e\tilde{w}^2}
		\left(\tilde{r}-\frac{\tilde{w}}{\sqrt{2}}\right)^2 + \tilde{z}^2\\
	(\tilde{U}_{1}^{1})_{ho} &=& \frac{32\tilde{v}}{e^2\tilde{w}^2}\left(\tilde{r}-\tilde{w}\right)^2+\tilde{z}^2,
	\end{array}
\end{eqnarray}
in the SHO approximation.  Defining~$(\tilde{\rho}_{p}^{l})_{ho} \equiv \rho \hoground$~for a 
specific~\lgmode{p}{l}~potential, the densities of states are
\begin{eqnarray}
	\label{tordensity}
	\begin{array}{lll}
		(\tilde{\rho}_0^1)_{ho} (\tilde{\epsilon}) &=& \sqrt{\frac{e}{\tilde{v}}} \frac{\tilde{w}^2}{12} 
			\tilde{\epsilon}^{3/2} \\
		(\tilde{\rho}_1^1)_{ho} (\tilde{\epsilon}) &=& \sqrt{\frac{e^2}{2\tilde{v}}}\frac{\tilde{w}^2}{12} 
			\tilde{\epsilon}^{3/2}.
	\end{array}
\end{eqnarray}
From here, we carry out the integrations in \refeq{npart} to find the BEC transition 
temperature,~$(T_{p}^{l})_{ho}$,
\begin{eqnarray}
	\label{tortc}
	\begin{array}{lll}
	(T_0^1)_c &=& 4 \left(\frac{\tilde{v}}{4\pi e}\right)^{1/5} \left(\frac{N}{\tilde{w}^2 \zeta(5/2)} 
		\right)^{2/5} \frac{\hoground}{k_B} \\
	(T_1^1)_c &=& 4 \left(\frac{\tilde{v}}{2\pi e^2}\right)^{1/5}\left(\frac{N}{\tilde{w}^2 \zeta(5/2)} 
		\right)^{2/5} \frac{\hoground}{k_B},
	\end{array}
\end{eqnarray}
where the term~$\zeta(5/2)$~is the Riemann Zeta function.\\
\indent Calculation of~$T_c$~using the full potential given by~\refeq{unitlesspot}~must 
be done numerically.  For a given~$\tilde{w}$~and~$\tilde{v}$, the 
density of states (\refeq{denofstates}) is numerically integrated for a
discrete set of energies, $\epsilon_i$.  The numerical integration is 
done in~\textit{Mathematica}~\cite{mathematica}~and we provide our source 
code online~\cite{lgnotebook}.  Twenty values of the 
density of states are calculated covering an energy range~$\epsilon_i = 0$~to~$\epsilon_{max}$.  
The value of~$\epsilon_{max}$~is determined 
such that most of the Bose-Einstein distribution is accounted 
for, while being certain that the energy range sufficiently 
maps out the most probable portions of the trap occupied by the particle.  To do this, we 
find the location of the critical temperature of the SHO approximation with respect to the potential 
barrier of the corresponding~\lgmode{p}{l}~mode.  The 
energy range is adjusted to span this region.
The limits of integration for~\refeq{denofstates}~are determined by 
solving~\refeq{unitlesspot}~for the classical volume accessible by a particle of energy~$\epsilon_i$.  
For the radial classical turning points, we use the Newton-Raphson numerical technique.  
The resulting set of points of the density of states,~$\rho_{i}(\epsilon_{i})$, are fit to the 
analytical expression, 
\begin{eqnarray}
	\label{model}
	\rho(\epsilon) = \sigma \epsilon^{\eta},
\end{eqnarray}
where the factors~$\sigma$~and~$\eta$~are the fitting parameters.  
This model has a general form that is characteristic for 
many systems.  For example, a Bose gas with no external potential 
results in~$\eta=1/2$, one confined by a 3-D isotropic oscillator gives 
$\eta=2$, and both harmonic approximations to the LG beams given in Equations 
(\ref{unitlesstor}) and (\ref{tordensity}) result in $\eta=3/2$.  We substitute 
the fitted function into~\refeq{npart}~and analytically integrate to find~$T_c$~for a 
given~$N$~(in our case, we choose a value of~$N=10^6$).  Finally, we analytically 
integrate~\refeq{energy}~to find the total energy,~$E(T)$, and then the heat capacity.  \\
%
%--------------------------------------------------------------------------
%
\begin{figure}
	\begin{center}
		\includegraphics[width=8.5cm]{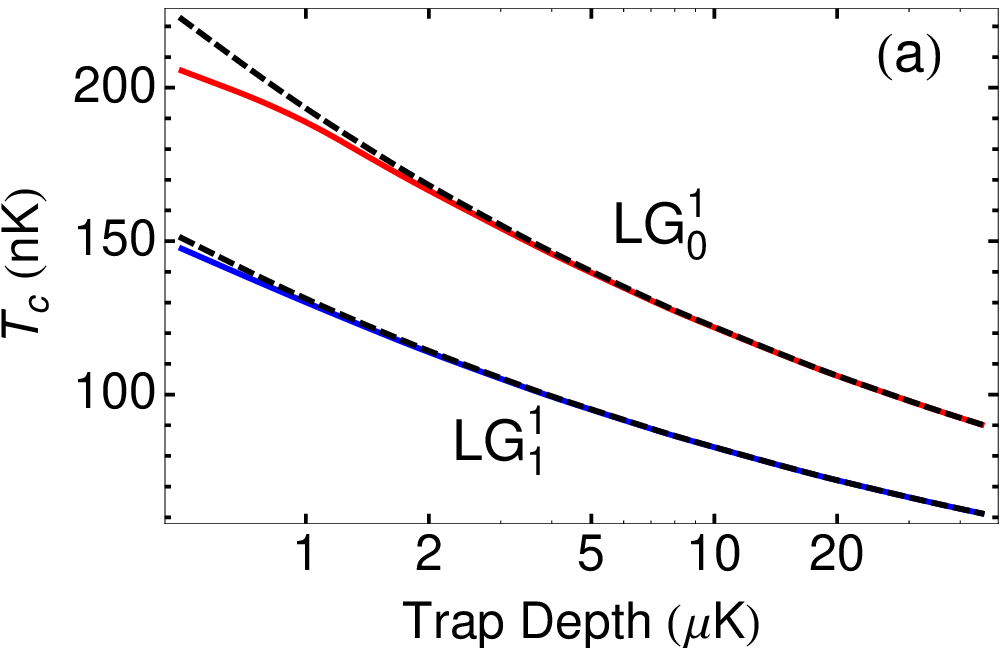}
		\includegraphics[width=8.5cm]{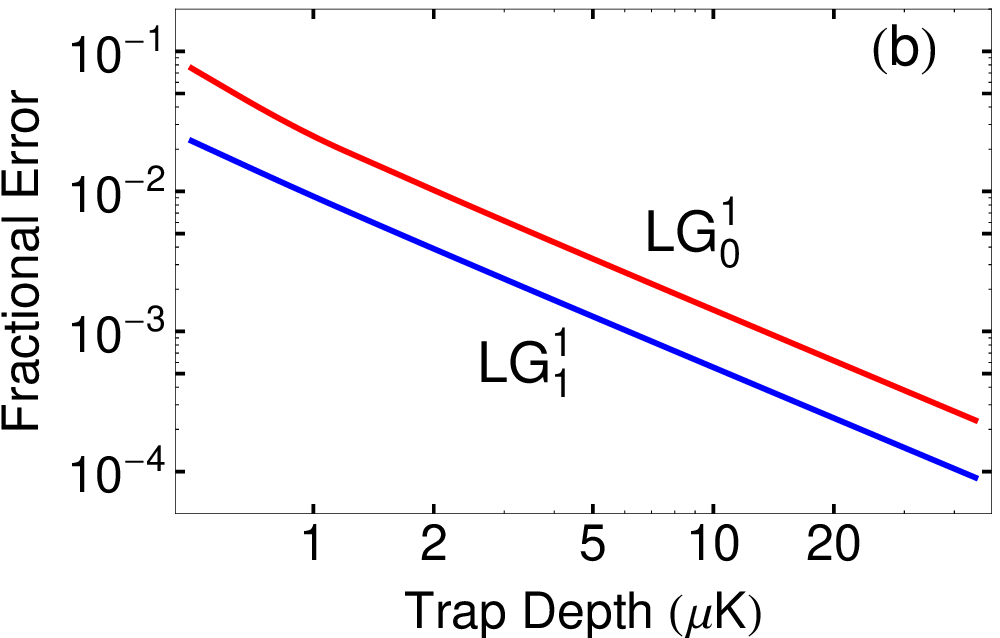}
		\caption{ (Color online) (a) $T_c$ 
		calculations using both the full potential (solid curve) and the SHO 
		approximation (dashed curve) as a function of trap depth, where the 
		radial trap frequency is held equal to the axial trap frequency, 
		$\omega_z/2\pi = 100 \textrm{Hz}$.  (b) The factional error of the 
		SHO calculation.}
		\label{tcsym}
	\end{center}
\end{figure}
\indent Figures~\ref{tcsym}~through~\ref{heatcapacity}~show the expected BEC 
transition temperature, ground state fraction, and heat capacity for a $^{87}$Rb 
gas confined in the~\lgmode{p}{l} modes.   We also show the corresponding 
harmonic approximations and compare with the exact calculation for different laser beam 
parameters.  \reffig{tcsym}~shows~$T_c$ in the~\lgmode{0}{1} (upper line) and the~\lgmode{1}{1} 
(lower line) modes as a function of the trap depth while maintaining equal trap 
frequencies in the axial and radial directions.  The trap depth is the 
height of the lowest confining barrier of the radial potential 
(\reffig{rpotential}).  The axial trap frequency is set to~$\omega_z/2\pi = 
100$~Hz, typical of magnetic trap experiments~\cite{dg99}.  
The trap depth is changed by adjusting~$\tilde{v}$, equivalent to 
adjusting the power or detuning of the trapping laser.  At the same time, $\tilde{w}$~is 
adjusted so the radial trap frequency is maintained at~$\omega_r = \omega_z$.  
The critical temperature is calculated using the full potential 
(solid line) and using the SHO approximation of the potential well (dotted line).  
Note that~$T_c$~decreases with the trap depth in this figure because of the 
constraint~$\omega_{r}=\omega_{z}$.  As the laser power increases, the beam waist 
must increase to maintain constant~$\omega_r$.  This increases the trap volume, and 
decreases~$T_c$.  \reffig{tcsym}~(b) shows the fractional error of the SHO 
approximations compared to the exact calculation.  The uncertainty is largest 
when the transition temperature approaches the trap depth.  The classical volume 
accessed by particles includes a larger fraction of the potential well where the 
harmonic approximation is no longer valid.  We calculate 
fractional differences from~$7.6\%$~to~$0.023\%$~over a trap depth 
of~$0.1~\mu\textrm{K}$~to~$10~\mu\textrm{K}$.\\
\indent In contrast to \reffig{tcsym}, \reffig{tcfreq} considers 
a symmetric case where the beam waist is held constant while increasing the trap depth.  
This corresponds to the trap volume decreasing as the trap depth increases.  
The critical temperature in the~\lgmode{0}{1}~(upper line) mode and the~\lgmode{1}{1}~(lower 
line) mode is plotted as a function of the laser power.  We fix the 
beam waist at~$w=42~\mu\textrm{m}$~ for the \lgmode{0}{1} mode, and $w=58~\mu\textrm{m}$ for 
the \lgmode{1}{1} mode.  The laser detuning is set to~$|\Delta|/2\pi = 1000$~GHz for both modes. 
The laser power is varied over a range of $0.52~\textrm{mW}-1.3~\textrm{W}$ for the \lgmode{0}{1} mode,
and $1.3~\textrm{mW}-5.9~\textrm{W}$.  This has the effect of increasing the trap depth and~$\omega_r$.  
To keep the trap symmetric, we allow $\omega_z$~to change such 
that $\omega_{z}=\omega_{r}$.  An increase in the trap depth has an effect of increasing~$T_c$.  
For this case, we also make the SHO approximation to the confining potential.  The fractional 
differences are from~$4.5\%$~to~$0.017\%$.  Like in \reffig{tcsym}, the largest error occurs when the trap 
depth and the transition temperature are comparable.  As a larger fraction of the particles populate 
larger regions of the potential, the particles occupy regions where the potential becomes less harmonic.  
We replace the constant, $\sigma$ in \refeq{model}, with the polynomial 
$\sigma(\epsilon) = \sigma_0 + \sigma_1 \epsilon + \sigma_2 \epsilon^2 \cdots$, where the minimum number 
of terms were kept such that the result converged. \\
\indent \reffig{tcasym} considers the effect on~$T_c$~when the trapping frequencies in the two trapping 
dimensions ($r$ and $z$) are allowed to vary with respect to each other.  
The symmetric case ($\omega_r/2\pi = 100$ Hz on the 
horizontal axis in~\reffig{tcasym}) corresponds to a laser power of~$P_0 = 10$~mW, 
a beam waist~$w = 50~\mu\textrm{m}$, and a detuning of~$|\Delta|/2\pi = 1000$~GHz.  
These values are concurrent with typical harmonic oscillator lengths 
of~$\holength = 1~\mu\textrm{m}$~reported by~\cite{dg99}.  We vary the 
radial trap frequency by fixing the waist of the trapping laser and increasing 
or decreasing the laser power.  In terms of the quantities defined in this paper, 
this is the same as keeping~$\tilde{w}$~fixed while increasing~$\tilde{v}$. 
Unlike the results in~\reffig{tcsym}, here the confinement in the~$z$-dimension 
is held constant leading to asymmetric traps.  We calculate~$T_c$~assuming the SHO approximations to the 
confining potential.  For radial frequencies~$\omega_r < \omega_z$, 
the atoms experience weaker confinement in the radial dimension, and the trapped atoms access regions 
of the trap where the potential is less harmonic.  The fractional error 
for the~\lgmode{0}{1}~(top line) and the~\lgmode{1}{1}~(bottom line) are 
plotted in~\reffig{tcasym}~(b). For a range 
of frequencies spanning~$40-400$~Hz, we find fractional differences of~$2.7-0.07\%$.\\ 
\indent We have also performed similar calculations for atoms trapped in a red detuned \lgmode{1}{1} laser beam.  
This confining potential is more complicated since it provides concentric multiply-connected traps.  These 
concentric traps have different trap depths and asymmetries.  However, the results are consistent with the 
calculations of the blue detuned \lgmode{1}{1} and red detuned \lgmode{0}{1} modes - BEC transition characteristics 
calculated from the full potential deviate from those using a SHO approximation when the classical volume of 
the potential accessed by particles when the system is near $T_c$ includes regions where the harmonic 
approximation to the potential breaks down. \\
\indent From~\refeq{npart}~the fraction of atoms in the ground state is
\begin{eqnarray}
	\label{eq:nground}
	\frac{N_0}{N} = 1-\frac{\int_0^{\infty}\frac{\rho(\epsilon)}{\textrm{Exp}(\beta\epsilon)-1}}{\int_0^{\infty}
		\frac{\rho(\epsilon)}{\textrm{Exp}(\beta_c\epsilon)-1}},
\end{eqnarray}
where $\beta_c=1/k_B T_c$.  \reffig{groundfrac}~(a) shows~$N_0/N$~as a function of the scaled 
temperature,~$T/T_c$~for the~\lgmode{0}{1}~mode dipole trap (dot-dashed line), 
the~\lgmode{1}{1}~mode dipole trap (solid line), the harmonic approximation 
to a~\lgmode{p}{l}~mode (dashed line), and the 3-D SHO (dotted line).  The trap parameters for the 
\lgmode{0}{1} mode consist of a laser power of $P_0 = 0.16$~mW 
and a beam waist of $w = 31~\mu$m.  The trap parameters for the \lgmode{1}{1} mode are $P_0 = 1.3$~mW 
and a beam waist of $w = 58$~mW.  Both modes have a laser detuning of $|\Delta|/2\pi = 1000$~GHz and have the 
same trapping frequency, $\omega_z = \omega_r = 2\pi \times 100$~Hz.  Note that given 
the form of the density of states (Equations (\ref{tordensity}) and (\ref{model})),~\refeq{eq:nground}~does 
not depend on the coefficient,~$\sigma$, but only on the exponent, $\eta$.  
Therefore,~$N_0/N$~is exactly the same for the SHO approximation to either potential. \\
\indent The 3-D SHO is included in~\reffig{groundfrac}~(a) 
to give a qualitative perspective.  The errors associated 
with making the harmonic approximation of the LG modes are similar to the 
differences between the multiply connected toroidal geometries and the 
singly-connected 3-D SHO potential. \reffig{groundfrac}~(b) reports the absolute 
difference between the two LG modes and their respective 
harmonic approximations.  The harmonic approximation 
error can reach as high as~$3.7\%$ for the~\lgmode{0}{1} and~$1.8\%$ for the~\lgmode{1}{1}.  
Such errors are on the same order as the effects associated with 
finite system sizes ($N\rightarrow\infty$~approximation) experimentally shown in 
reference~\cite{ejm96}.   The trap depth is~$0.5~\mu\textrm{K}$, the beam waist for the 
\lgmode{0}{1} is~$31~\mu\textrm{m}$, the beams waist for the~\lgmode{1}{1} is~$58~\mu\textrm{m}$, 
and the laser detuning is $|\Delta|/2\pi=1000$~GHz.   A precise knowledge of the fraction of 
trapped atoms in the ground state may be necessary for many of the applications of LG laser modes as dipole traps.  \\
\indent \reffig{heatcapacity} shows the heat capacity for samples 
trapped in~\lgmode{0}{1}~(upper curve) and~\lgmode{1}{1}~(middle curve) 
with a respective harmonic approximation (lower dashed curve).  Like~\reffig{groundfrac}, the harmonic 
approximation can be represented by a single curve because the scaling of the heat capacity 
and temperature allow for a dependence on a single parameter,~$\eta$~in~\refeq{model}. The 
laser parameters correspond to situations where the confined atoms occupy regions of the 
trap that deviate from a harmonic description. The disagreement between the LG potentials 
and the harmonic approximations can be as large as~$24\%$~around~$T_c$. \\
%
%--------------------------------------------------------------------------
%
\indent Toroidal traps are ideal geometries for employment in BEC gyroscope and vortex experiments.  
Matter-wave interferometers may offer better sensitivity than traditional optical interferometers.  
Vortices provide a key testing ground for superfluid behavior studies.  For these precise applications, 
BEC characteristics should be well understood.  In particular, knowledge of the fraction of trapped 
atoms in the condensate is important for these applications.  In addition, the wave functions of 
condensates confined in LG traps may need to take into account the full trap geometry. \\
\indent We calculate the thermal properties of 
a Bose gas confined by a~\lgmode{p}{l}~laser dipole trap: the critical 
transition temperature for a wide range of experimental parameters, the 
condensate fraction, and the heat capacity.  We also compute the thermal 
properties using a SHO approximation to the potential minimum and compare 
to the exact results.  Depending on the precision required, there exists a regime in which 
the thermal properties of atoms confined to~\lgmode{p}{l}~dipole traps contain non-negligible deviations.  
When the critical temperature is on the order of the trap depth (a depth of $0.1~\mu\textrm{K}$), 
it is predicted to be too high by errors as large as~$8\%$.  In this regime, the number of atoms in 
the ground state is underestimated for temperatures below the critical temperature.  We also find that the heat 
capacity contains large deviations around the BEC transition temperature.  
These corrections are on the order of well known effects (such as finite size corrections). \\
\indent This work is supported by the Research Corporation, Digital Optics 
Corporation, and The University of Oklahoma.  
%
%
%--------------------------------------------------------------------------
%	BibTex References
%--------------------------------------------------------------------------
%
%
\bibliographystyle{elsarticle-num}
\bibliography{bib_lgbec}
%-------------------------------------------------------------------------
%
%
\end{document}